\documentclass[aps,pra,reprint,nobibnotes]{revtex4-1}
\usepackage{graphicx}
\usepackage{dcolumn}
\usepackage{amsmath}
\newcommand{\bfr}{{\bf r}}
\newcommand{\ud}{\mathrm{d}}

\begin{document}

\title{Dispersion interactions from a local polarizability model}
\author{Oleg A. Vydrov}
\author{Troy Van Voorhis}
\affiliation{Department of Chemistry, Massachusetts Institute of Technology,
 Cambridge, MA, 02139, USA}

\date{\today}

\begin{abstract}
A local approximation for dynamic polarizability leads to a nonlocal functional for
the long-range dispersion interaction energy via an imaginary-frequency integral.
We analyze several local polarizability approximations and argue that the form
underlying the construction of our recent van der Waals functional
[O.~A. Vydrov and T. Van Voorhis, Phys. Rev. Lett. \textbf{103}, 063004 (2009)]
is particularly well physically justified. Using this improved formula, we compute
dynamic dipole polarizabilities and van der Waals $C_6$ coefficients for a set of
atoms and molecules. Good agreement with the benchmark values is obtained in most cases.
\end{abstract}
\maketitle

\section{Introduction}

Recently we developed \cite{VV09} a nonlocal correlation energy functional that describes
the entire range of van der Waals interactions in a general and seamless fashion, using only
the electron density and its gradient as input. Improving upon its predecessors \cite{vdW-DF-04,
vdW-DF-09}, the new van der Waals density functional \cite{VV09}, denoted VV09, has a simple
analytic form, generalized to spin-polarized systems and well-behaved in some important limits.
In the asymptotic long-range regime, VV09 reduces to a form similar to the models of
Refs.~\cite{ALL-96} and \cite{Dobson-96}, yet with some crucial differences. In this article,
we examine this long-range behavior in detail and present some test results of dynamic dipole
polarizabilities and asymptotic van der Waals $C_6$ coefficients.

\section{Formalism}

For two compact systems $A$ and $B$ separated by a large distance $R$, the nonretarded dispersion
interaction energy \cite{Kaplan} behaves asymptotically as $-C_6^{AB} R^{-6}$ with the $C_6$
coefficient given by the formula \cite{Stephen}
\begin{equation}
 C^{AB}_6 = \frac{3\hbar}{\pi} \int_0^\infty \! \ud u \,
 \bar{\alpha}^A(iu) \, \bar{\alpha}^B(iu),
 \label{C6}
\end{equation}
where $\bar{\alpha}(iu)$ is the average (isotropic) dynamic dipole polarizability at imaginary frequency $iu$.
A simple but often sufficiently accurate approximation is to describe $\bar{\alpha}$ by a local model:
\begin{equation}
 \bar{\alpha}(iu) = \int \ud\bfr\, \alpha(\bfr,iu).
 \label{alpha-bar}
\end{equation}
The long-range dispersion interaction energy between systems $A$ and $B$ can then be written
\cite{Book-98,Dobson-98} in terms of local polarizabilities as
\begin{equation}
 E_\text{disp} = -\frac{3\hbar}{\pi} \int_0^\infty \! \ud u \int_A \ud\bfr \int_B \ud\bfr'\,
   \frac{\alpha(\bfr,iu) \, \alpha(\bfr',iu)}{\left|\bfr-\bfr'\right|^6},
 \label{E-disp}
\end{equation}
where $\bfr$ is within the domain of system $A$ and $\bfr'$ is within the domain of $B$.

In Refs.~\cite{ALL-96,Dobson-96,Dobson-98,ALL-98}, a simple model for $\alpha(\bfr,iu)$ was
derived from the response properties of a uniform electron gas (UEG). The zero wave vector
UEG dielectric function at frequency $\omega$ is given by
\begin{equation}
 \epsilon(\omega) = 1 - \frac{\omega_p^2}{\omega^2},
 \label{epsilon-UEG}
\end{equation}
where $\omega_p = \sqrt{4\pi n e^2/m}$ is the plasma frequency for the electron density $n$.
In nonuniform systems, the local analog of $\omega_p$ can be defined via $\omega_p^2(\bfr) =
4\pi n(\bfr) e^2/m$. Then the local polarizability for $\omega = iu$ is found as
\cite{ALL-96,ALL-98}
\begin{equation}
 \alpha(\bfr,iu) = \frac{1}{4\pi}\!\left[1 - \frac{1}{\epsilon(\bfr,iu)}\right]
  = \frac{1}{4\pi} \, \frac{\omega_p^2(\bfr)}{\omega_p^2(\bfr)+u^2}.
  \label{alpha-ALL}
\end{equation}
Plugging Eq.~(\ref{alpha-ALL}) into Eq.~(\ref{E-disp}) we arrive at the
Andersson--Langreth--Lundqvist (ALL) formula \cite{ALL-96}
\begin{equation}
 E_\text{disp} = -\frac{3\hbar}{32\pi^2} \int_A \ud\bfr \int_B \ud\bfr'
  \frac{\omega_p(\bfr) \omega_p(\bfr')}{\omega_p(\bfr) + \omega_p(\bfr')}
  \left|\bfr-\bfr'\right|^{-6}.
 \label{E-ALL}
\end{equation}
An immediately apparent problem with Eq.~(\ref{alpha-ALL}) is its treatment of static
polarizability:
\begin{equation}
 \bar{\alpha}(0) = \int \ud\bfr\, \alpha(\bfr,0) = \int \ud\bfr\, \frac{1}{4\pi}.
\end{equation}
$\alpha(\bfr,0)$ is constant everywhere, therefore the above integral is divergent unless a
cutoff is introduced. Eq.~(\ref{E-ALL}), taken as it is, yields finite but severely
overestimated $E_\text{disp}$. These difficulties are circumvented \cite{Rapcewicz,ALL-96,ALL-98}
by the introduction of sharp density-based integration cutoffs in Eqs.~(\ref{alpha-bar}) and
(\ref{E-ALL}). Calculated polarizabilities and $C_6$ coefficients are admittedly \cite{ALL-96,ALL-98}
sensitive to the choice of the cutoff criterion, although the prescription of
Refs.~\cite{Rapcewicz,ALL-96,ALL-98} appears to work well in many cases. Note that Ref.~\cite{ALL-98}
gave separate cutoff criteria for the spin-compensated and the fully spin-polarized cases. To our
knowledge, a prescription for a general spin-polarization case has never been put forth.

An integration cutoff discards density tail regions, which is not entirely satisfactory from the
formal point of view. In the $u \to \infty$ limit, the $f$-sum rule requires \cite{Mahan-book} that
\begin{equation}
 \bar{\alpha}(iu) \to \frac{N e^2}{m u^2} = \int\!\ud\bfr\,\frac{\omega_p^2(\bfr)}{4\pi u^2},
\end{equation}
where $N$ is the number of electrons in the system. Omission of the density tails leads to the
reduction of $N$, in violation of the $f$-sum rule.

Another formal shortcoming of the ALL theory was pointed out by Nesbet \cite{Nesbet-97,Nesbet-98},
who argued that a more appropriate relation between $\alpha$ and $\epsilon$ is given by the
Clausius--Mossotti formula
\begin{equation}
 \alpha = \frac{3}{4\pi} \, \frac{\epsilon-1}{\epsilon+2},
 \label{Clausius}
\end{equation}
so that Eq.~(\ref{alpha-ALL}) should be replaced by
\begin{equation}
 \alpha(\bfr,iu) = \frac{1}{4\pi} \, \frac{\omega_p^2(\bfr)}{\omega_p^2(\bfr)/3 + u^2}.
 \label{alpha-Nesbet}
\end{equation}
It appears that Nesbet's articles went unnoticed, because in the numerous practical
applications \cite{Andersson-99,Hirao-02,Hirao-05a,Hirao-05b,Hirao-07,Hirao-08,Hirao-09a,Hirao-09b,Hirao-10,
Silvestrelli-08,Silvestrelli-09a,Silvestrelli-09b,Silvestrelli-09c,Silvestrelli-09d,Cremer-09,Hagiwara-09,
Rotenberg-10} of the ALL formula, Nesbet's suggestion was never utilized.

The validity of Eq.~(\ref{alpha-Nesbet}) is corroborated by the example of interacting
jellium spheres. For two identical spheres of uniform density and radius $r_0$ separated by
distance $R$ (such that $R \gg r_0$) the interaction energy is given by \cite{spheres}
\begin{equation}
 E_\text{spheres} = - \frac{\sqrt{3}}{4} \, \hbar \omega_p \frac{r_0^6}{R^6}.
 \label{spheres}
\end{equation}
The above result is exactly reproduced if Nesbet's model of Eq.~(\ref{alpha-Nesbet}) is plugged
into Eq.~(\ref{E-disp}), whereas the ALL formula (\ref{E-ALL}) underestimates this result by the
factor of $3\sqrt{3} \approx 5$. We note in passing that all three versions of the vdW-DF functional
of Refs.~\cite{vdW-DF-04,vdW-DF-09,vdW-DF-10} fail to reproduce Eq.~(\ref{spheres}) even on the
qualitative level, yielding incorrect dependence on the electron density.

Local polarizability models of Eqs.~(\ref{alpha-ALL}) and (\ref{alpha-Nesbet}) were derived
using the UEG dielectric function of Eq.~(\ref{epsilon-UEG}). UEG is rather dissimilar to
our target systems --- molecules. UEG has a continuous excitation spectrum and a zero band
gap (i.e.\ it is a metal), whereas molecules have a discrete spectrum with a gap between
the ground state and the fist excited state. The polarizability model could be made
more realistic by introducing a gap. For a semiconductor with a band gap $\hbar \omega_g$,
the zero wave vector dielectric function \cite{Levine} is typically written as
\begin{equation}
 \epsilon(\omega) = 1 + \frac{\omega_p^2}{\omega_g^2 - \omega^2}.
 \label{epsilon_gap}
\end{equation}
Using this $\epsilon(\omega)$ in the Clausius--Mossotti formula (\ref{Clausius}), we obtain
\begin{equation}
 \alpha(\bfr,iu) = \frac{1}{4\pi} \, \frac{\omega_p^2(\bfr)}{\omega_p^2(\bfr)/3
 + \omega_g^2(\bfr) + u^2},
 \label{my-alpha}
\end{equation}
where we introduced a ``local gap'' $\hbar \omega_g(\bfr)$. The above $\alpha(\bfr,iu)$
leads via Eq.~(\ref{E-disp}) to the energy expression
\begin{equation}
 E_\text{disp} = -\frac{3\hbar}{32\pi^2}\! \int_A \ud\bfr \int_B \ud\bfr'
    \frac{\omega^2_p(\bfr) \, \omega^2_p(\bfr') \, \left|\bfr-\bfr'\right|^{-6}}
    {\omega_0(\bfr) \omega_0(\bfr') \big[\omega_0(\bfr) + \omega_0(\bfr')\big]},
 \label{my-E}
\end{equation}
where $\omega_0 = \sqrt{\omega_g^2 + \omega_p^2/3}$. A suitably chosen $\omega_g(\bfr)$
obviates any need for an integration cutoff in $E_\text{disp}$ and $\bar{\alpha}(iu)$.
As a result, the $f$-sum rule on $\bar{\alpha}(iu)$ is obeyed.

An apt model for $\omega_g(\bfr)$ can be deduced by examining the behavior of the electron
density $n(\bfr)$. In atoms, $n(\bfr)$ can be approximated as piecewise exponential.
In the density tails, the exact behavior \cite{Levy-84} is known:
\begin{equation}
 n(\bfr) \sim \exp\left(-\alpha |\bfr|\right), \quad
 \text{with} \quad \alpha = 2 \left(2m I / \hbar^2 \right)^{1/2},
 \label{tail}
\end{equation}
where $I$ is the ionization potential. Generalizing the result of Eq.~(\ref{tail}),
we can define a ``local ionization potential'' as \cite{Savin-99,Savin-07}
\begin{equation}
 I(\bfr) = \frac{\hbar^2}{8 \, m} \left| \frac{\nabla n(\bfr)}{n(\bfr)} \right|^2.
\end{equation}
Taking $\hbar \omega_g(\bfr) \propto I(\bfr)$, in Ref.~\cite{VV09} we defined
\begin{equation}
 \omega_g^2(\bfr) = C \frac{\hbar^2}{m^2}
 \left| \frac{\nabla n(\bfr)}{n(\bfr)} \right|^4,
 \label{gap}
\end{equation}
where $C$ is an adjustable parameter. We fitted $C$ to a benchmark set of 17 van der Waals $C_6$
coefficients and obtained \cite{VV09} the optimal value of $C = 0.0089$. It is instructive to
consider the ratio $\omega_g (\bfr) / I (\bfr) = 8 \sqrt{0.0089} = 0.755$. This ratio seems
reasonable since $\omega_g$ should be somewhat smaller than $I$. In the uniform density limit,
Eq.~(\ref{gap}) gives $\omega_g = 0$, so that our $\alpha(\bfr,iu)$ of Eq.~(\ref{my-alpha})
reduces to Nesbet's $\alpha(\bfr,iu)$ of Eq.~(\ref{alpha-Nesbet}).

\begin{figure*}[tbp!]
\includegraphics[width=0.75\textwidth]{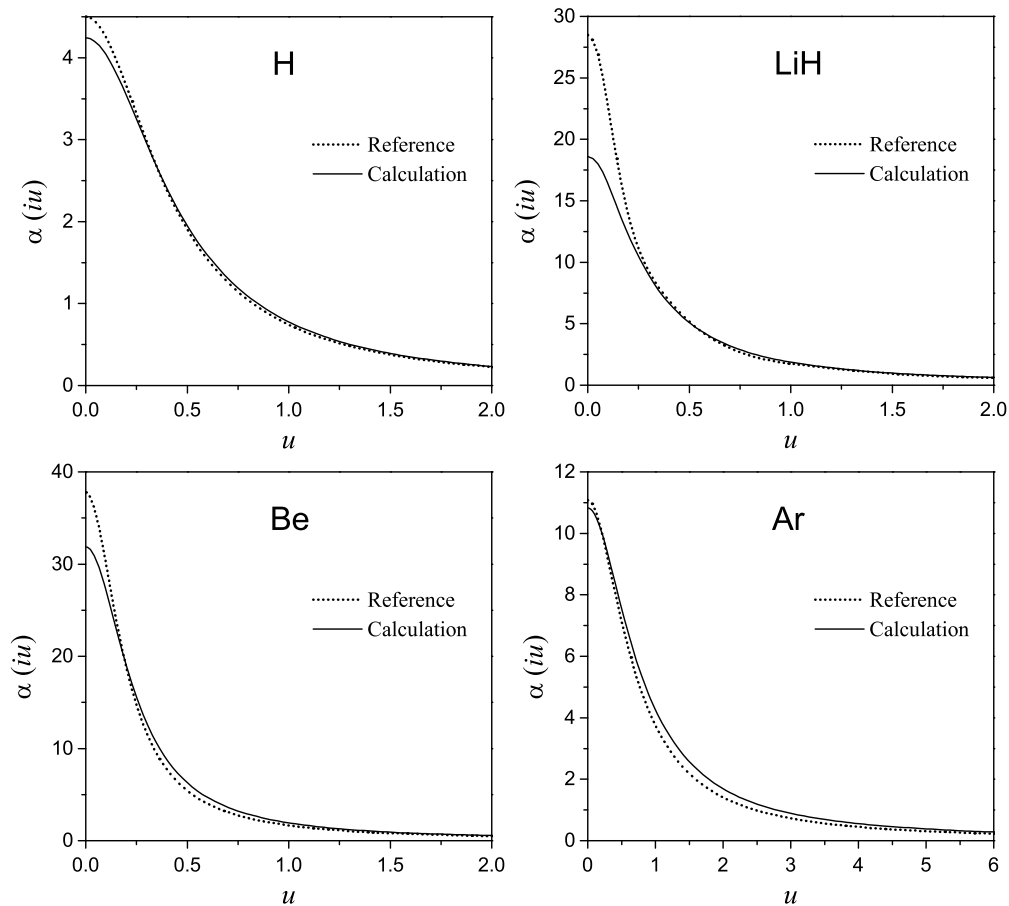}
\caption{Average dynamic dipole polarizabilities at imaginary frequencies calculated using
 the VV09 model, i.e.\ via Eqs.~(\ref{alpha-bar}) and (\ref{my-alpha}). Atomic units are used.
 The reference values are from Ref.~\cite{LiH} for LiH and from Ref.~\cite{Derevianko-10} for the atoms.}
 \label{fig-alpha}
\end{figure*}

Eqs.~(\ref{my-alpha}), (\ref{my-E}), and (\ref{gap}) require only the \textit{total} electron
density as input and include no dependence on spin-polarization. The question of the proper
treatment of spin, not fully resolved in the ALL theory \cite{ALL-98}, does not arise in this model.
We mention in passing that vdW-DF functionals of Refs.~\cite{vdW-DF-04,vdW-DF-09,vdW-DF-10}
were defined only for the spin-compensated case and their extension to spin-polarized systems
is nontrivial. In VV09 \cite{VV09}, the dependence on spin-polarization enters only at shorter
range. In the long-range limit, the VV09 nonlocal correlation energy reduces to Eq.~(\ref{my-E}).
In this regard, a clarification should be made: The coefficient before the
double integral in Eq.~(\ref{my-E}) is twice the coefficient in Eq.~(13) of Ref.~\cite{VV09}
because these formulas compute different things. Eq.~(\ref{my-E}) computes the interaction
energy between systems $A$ and $B$, hence the integral over $\bfr$ is limited to the part of
space confining system $A$, while the integral over $\bfr'$ is limited to the domain of $B$.
On the other hand, Eq.~(13) of Ref.~\cite{VV09} gives the nonlocal correlation energy, which
includes inter- and intramolecular contributions, hence both $\bfr$ and $\bfr'$ integrals
are over the entire space.

\begin{table}[tbp!]
\caption{Isotropic polarizabilities $\bar{\alpha}(iu)$ calculated via Eqs.~(\ref{alpha-bar})
 and (\ref{my-alpha}) compared to the reference values for BeH$_2$ \cite{BeH2} and BH \cite{BH}.
 Atomic units are used.
\label{tab-alpha}}
\begin{ruledtabular}
\begin{tabular}{ldddd}
 & \multicolumn{2}{c}{BeH$_2$} & \multicolumn{2}{c}{BH} \\ \cline{2-3}\cline{4-5}
 $u$ & \multicolumn{1}{c}{Ref.} & \multicolumn{1}{c}{Calc.}
 & \multicolumn{1}{c}{Ref.} & \multicolumn{1}{c}{Calc.} \\ \hline
0.0	 & 19.760 & 19.413 & 21.430 & 17.442 \\
0.142857 & 17.234 & 17.049 & 16.582 & 15.506 \\
0.333333 & 11.280 & 11.509 & 10.343 & 10.738 \\
0.6	 &  6.084 &  6.440 &  5.625 &  6.139 \\
1.0	 &  2.917 &  3.161 &  2.755 &  3.044 \\
1.666667 &  1.233 &  1.360 &  1.198 &  1.307 \\
3.0	 &  0.418 &  0.480 &  0.424 &  0.453 \\
7.0	 &  0.081 &  0.105 &  0.091 &  0.096 \\
\end{tabular}
\end{ruledtabular}
\end{table}

\section{Benchmark tests}

All calculations reported in this section were performed at the LC-$\omega$PBE08 \cite{LC-wPBE08}
electron densities (using $\omega$ = 0.45~$a_0^{-1}$, as suggested in Ref.~\cite{LC-wPBE08}),
except for the H atom polarizability, computed at the Hartree-Fock (i.e.\ exact in this case)
density. For the numerical integration, we use the Euler-Maclaurin-Lebedev unpruned (75,302)
quadrature grid. The aug-cc-pVQZ basis set is used in all calculations. All the numbers in this
section are given in atomic units (a.u.).

Using the VV09 model, given by Eqs.~(\ref{my-alpha}) and (\ref{gap}), we have calculated the
isotropic dynamic dipole polarizabilities as functions of imaginary frequencies for several
atoms and small molecules for which accurate reference data \cite{Derevianko-10,LiH,BeH2,BH}
are available. For LiH, BeH$_2$, and BH, we used the same bond lengths as in
Refs.~\cite{LiH,BeH2,BH}. The results are given in Fig.~\ref{fig-alpha} and
Table~\ref{tab-alpha}. The agreement between the calculated and reference values of
$\bar{\alpha}(iu)$ is generally quite good, although this method has a tendency of
underestimating static polarizabilities $\bar{\alpha}(0)$. The largest errors in
$\bar{\alpha}(0)$ are observed for LiH (Fig.~\ref{fig-alpha}) and for alkali-metal
atoms (not shown). Underestimation of $\bar{\alpha}(0)$ causes rather large errors
in $C_6$ coefficients for alkali-metal atoms, as shown below.

\begin{table*}[tbp!]
\caption{$C_6^{AA}$ coefficients (a.u.)\ for closed-shell species calculated by several methods.
 Experimental geometries \cite{CRC} are used for all molecules.
 MPE stands for the mean percentage error and MAPE stands for the mean absolute percentage error.
\label{tab-C6-big}}
\begin{ruledtabular}
\begin{tabular}{ldddddc}
Molecule & \multicolumn{1}{c}{vdW-DF-04\footnotemark[1]} &
 \multicolumn{1}{c}{vdW-DF-09\footnotemark[2]} & \multicolumn{1}{c}{vdW-DF-10\footnotemark[3]}
 & \multicolumn{1}{c}{VV09\footnotemark[4]} & \multicolumn{1}{c}{Accurate} & Ref.\footnotemark[5] \\ \hline
He            & 2.93  & 1.63  & 0.76  & 1.45  & 1.46  & \cite{Derevianko-10}     \\
Ne            & 9.45  & 6.52  & 3.07  & 8.44  & 6.35  & \cite{Kumar-10}          \\
Ar            & 62.67 & 61.41 & 25.29 & 70.08 & 64.42 & \cite{Kumar-10}          \\
Kr            & 114.3 & 120.0 & 47.7  & 131.2 & 130.1 & \cite{Kumar-10}          \\
Be            & 269   & 330   & 102   & 186   & 214   & \cite{Derevianko-10}     \\
Mg            & 649   & 835   & 246   & 425   & 627   & \cite{Derevianko-10}     \\
Zn            & 269   & 240   & 87    & 163   & 284   & \cite{Chu-JCP-04}        \\
H$_2$         & 16.82 & 12.53 & 5.09  & 10.28 & 12.09 & \cite{Meath-90-IntJQC}   \\
N$_2$         & 78.76 & 77.59 & 31.96 & 88.70 & 73.43 & \cite{Meath-90-IntJQC}   \\
Cl$_2$        & 289.3 & 336.8 & 131.4 & 366.7 & 389.2 & \cite{Meath-02-MolPhys}  \\
HF            & 23.12 & 18.01 & 7.97  & 21.13 & 19.00 & \cite{Meath-85-MolPhys}  \\
HCl           & 114.3 & 119.9 & 47.2  & 124.6 & 130.4 & \cite{Meath-85-MolPhys}  \\
HBr           & 180.1 & 198.2 & 76.1  & 200.2 & 216.6 & \cite{Meath-85-MolPhys}  \\
CO            & 87.56 & 86.34 & 35.01 & 93.51 & 81.40 & \cite{Meath-82-CP}       \\
CO$_2$        & 127.6 & 130.6 & 54.5  & 159.4 & 158.7 & \cite{Meath-82-CP}       \\
CS$_2$        & 586.3 & 731.7 & 274.3 & 739.4 & 871.1 & \cite{Meath-84-CP}       \\
OCS           & 316.8 & 370.1 & 143.4 & 395.6 & 402.2 & \cite{Meath-84-CP}       \\
N$_2$O        & 136.1 & 140.3 & 58.4  & 172.4 & 184.9 & \cite{Meath-78-JCP}      \\
CH$_4$        & 122.0 & 130.1 & 50.8  & 129.6 & 129.6 & \cite{Meath-80-CP}       \\
CCl$_4$       & 1436  & 1882  & 715   & 2044  & 2024  & \cite{Meath-02-THEO}     \\
NH$_3$        & 82.47 & 79.32 & 32.00 & 82.78 & 89.03 & \cite{Meath-78-JCP}      \\
H$_2$O        & 46.96 & 40.83 & 17.17 & 44.95 & 45.29 & \cite{Meath-78-JCP}      \\
SiH$_4$       & 338.0 & 406.1 & 147.2 & 344.6 & 343.9 & \cite{Meath-03-CP}       \\
SiF$_4$       & 360.9 & 382.7 & 158.6 & 455.8 & 330.2 & \cite{Meath-03-MolPhys}  \\
H$_2$S        & 186.6 & 208.9 & 79.1  & 200.3 & 216.8 & \cite{Meath-88-CanJChem} \\
SO$_2$        & 239.5 & 265.1 & 106.5 & 305.2 & 294.0 & \cite{Meath-84-CP}       \\
SF$_6$        & 568.0 & 659.7 & 274.6 & 869.9 & 585.8 & \cite{Meath-85-JCP}      \\
C$_2$H$_2$    & 191.3 & 210.3 & 81.0  & 210.3 & 204.1 & \cite{Meath-92-MolPhys}  \\
C$_2$H$_4$    & 259.7 & 293.8 & 113.2 & 297.3 & 300.2 & \cite{Meath-07-CanJChem} \\
C$_2$H$_6$    & 330.4 & 386.1 & 148.8 & 396.6 & 381.8 & \cite{Meath-80-CP}       \\
CH$_3$OH      & 194.0 & 208.5 & 83.2  & 226.1 & 222.0 & \cite{Meath-05-CzechCC}  \\
CH$_3$OCH$_3$ & 458.7 & 532.5 & 207.7 & 567.9 & 534.0 & \cite{Meath-08-MolPhys}  \\
Cyclopropane  & 480.7 & 596.1 & 228.3 & 632.6 & 630.8 & \cite{Meath-80-CP}       \\
C$_6$H$_6$    & 1297  & 1715  & 647   & 1838  & 1723  & \cite{Meath-92-MolPhys}  \\ \hline
MPE (\%)      &  -2.8 &  -0.5 & -60.9 &   1.2 &       & \\
MAPE (\%)     &  18.5 &  10.4 &  60.9 &  10.7 &       & \\
\end{tabular}
\end{ruledtabular}
\footnotetext[1]{The method of Ref.~\cite{vdW-DF-04}.}
\footnotetext[2]{The method of Ref.~\cite{vdW-DF-09}.}
\footnotetext[3]{This method is denoted as vdW-DF2 in Ref.~\cite{vdW-DF-10}.}
\footnotetext[4]{The formalism proposed in Ref.~\cite{VV09} and described in this work.}
\footnotetext[5]{Literature references for the accurate benchmark $C_6^{AA}$ values.} 
\end{table*} 

When the distance between species $A$ and $B$ is large compared to the size of these systems,
$|\bfr-\bfr'|^{-6}$ in Eq.~(\ref{my-E}) can be taken out of the integral as $R^{-6}$, leading
to the $-C_6^{AB} R^{-6}$ form with $C_6^{AB}$ given by Eq.~(\ref{C6}). To further assess the
quality of the VV09 local polarizability model, we have calculated isotropic dispersion $C_6$
coefficients for a number of atoms and molecules. As expected from Eq.~(\ref{C6}),
any errors in the polarizability $\bar{\alpha}^A(iu)$ are reflected in $C_6^{AA}$ and similarly
in $C_6^{AB}$. It is sufficient to include only $C_6^{AA}$ in our benchmark set, since the
accuracy for $C_6^{AA}$ and $C_6^{BB}$ determines the accuracy for $C_6^{AB}$. For example,
VV09 strongly underestimates the $C_6$ coefficient for the Li--Li interaction, and as a
result, all $C_6$ coefficients for Li interacting with other species are also underestimated.
On the contrary, VV09 gives very accurate $C_6$ coefficients for He--He and Kr--Kr, and
consequently, $C_6$ for He--Kr is also very accurate.

In Ref.~\cite{VV09} we reported the $C_6^{AA}$ coefficients for a set of 17 closed-shell species,
computed within the VV09 methodology. In fact, the value of $C = 0.0089$ in Eq.~(\ref{gap}) was
fitted to that set. In this study, we test whether this fit is transferable to atoms and molecules
outside of the training set. In Table~\ref{tab-C6-big} we assembled a set of 34 closed-shell
species for which accurate $C_6^{AA}$ are known. Using this benchmark set, we compare the accuracy
of VV09 \cite{VV09} to the similar methods of Refs.~\cite{vdW-DF-04,vdW-DF-09,vdW-DF-10}.
In the asymptotic limit, all these methods reduce to the form of Eq.~(\ref{my-E}), but with
different models for $\omega_0$, as discussed in Ref.~\cite{vdW-DF-09}.
Deviations from the reference values are summarized in Table~\ref{tab-C6-big}
as mean (signed) percentage errors (MPE) and mean absolute percentage errors (MAPE). VV09 and
vdW-DF-09 exhibit very similar accuracy with MAPE of just over 10\%. vdW-DF-04 is somewhat
less accurate with MAPE of 18.5\%. The latest reparameterization (denoted as vdW-DF2 in
Ref.~\cite{vdW-DF-10}, but called vdW-DF-10 here for consistency) yields very poor $C_6$
coefficients: as compared to the reference values, vdW-DF-10 underestimates $C_6^{AA}$ by
a factor of 2.6 on average.

The good performance of vdW-DF-09 for $C_6$ coefficients motivated Sato and Nakai \cite{Nakai}
to devise a pairwise atom-atom dispersion correction using the local polarizability model
\cite{vdW-DF-09} underlying the construction of vdW-DF-09. We believe that the VV09 model of
Eq.~(\ref{my-alpha}) can also be successfully employed in this scheme.

As mentioned above, none of the three versions of vdW-DF \cite{vdW-DF-04,vdW-DF-09,vdW-DF-10}
has been generalized for open-shell systems, whereas VV09 is defined for a general spin-polarized
case. In Table~\ref{tab-C6-open}, we compare the $C_6^{AA}$ coefficients predicted by VV09 to
the accurate reference values for 20 open-shell species. The agreement is satisfactory in most cases.
The largest errors are observed for the alkali-metal atoms Li and Na. The strong underestimation
of the $C_6$ coefficients for alkali-metal atoms was also noted for the ALL formula \cite{ALL-96}.
It is likely that the local approximation of Eq.~(\ref{alpha-bar}) is inadequate for such highly
polarizable systems as alkali metals.

\begin{table}[tbp!]
\caption{$C_6^{AA}$ coefficients (a.u.)\ for open-shell species calculated using Eq.~(\ref{my-E}).
 \label{tab-C6-open}}
\begin{ruledtabular}
\begin{tabular}{lddc}
Molecule & \multicolumn{1}{c}{VV09} & \multicolumn{1}{c}{Accurate} & Ref. \\ \hline
H     & 6.75  & 6.50  & \cite{Derevianko-10} \\
Li    & 565   & 1389  & \cite{Derevianko-10} \\
B     & 87.6  & 99.5  & \cite{Chu-JCP-04} \\
C     & 47.0  & 46.6  & \cite{Chu-JCP-04} \\
N     & 27.65 & 24.10 & \cite{Meath-78-JCP} \\
O     & 18.19 & 14.89 & \cite{Meath-78-JCP} \\
F     & 12.21 & 9.52  & \cite{Chu-JCP-04} \\
Na    & 669   & 1556  & \cite{Derevianko-10} \\
Al    & 353   & 528   & \cite{Chu-JCP-04} \\
Si    & 253   & 305   & \cite{Chu-JCP-04} \\
P     & 179   & 185   & \cite{Chu-JCP-04} \\
S     & 130   & 134   & \cite{Chu-JCP-04} \\
Cl    & 94.7  & 94.6  & \cite{Chu-JCP-04} \\
Ga    & 255   & 498   & \cite{Chu-JCP-04} \\
Ge    & 251   & 354   & \cite{Chu-JCP-04} \\
As    & 222   & 246   & \cite{Chu-JCP-04} \\
Se    & 190   & 210   & \cite{Chu-JCP-04} \\
Br    & 158   & 162   & \cite{Chu-JCP-04} \\
O$_2$ & 66.18 & 61.57 & \cite{Meath-78-JCP} \\
NO    & 77.83 & 69.73 & \cite{Meath-78-JCP} \\ \hline
MPE (\%)  & -9.8 & & \\
MAPE (\%) & 18.7 & & \\
\end{tabular}
\end{ruledtabular}
\end{table} 

\section{Conclusions}

The ALL formula (\ref{E-ALL}) for the long-range dispersion energy enjoys growing popularity
\cite{Andersson-99,Hirao-02,Hirao-05a,Hirao-05b,Hirao-07,Hirao-08,Hirao-09a,Hirao-09b,Hirao-10,
Silvestrelli-08,Silvestrelli-09a,Silvestrelli-09b,Silvestrelli-09c,Silvestrelli-09d,Cremer-09,
Hagiwara-09,Rotenberg-10}, even though it has been superseded by more general
\cite{VV09,vdW-DF-04,vdW-DF-09,vdW-DF-10} and more accurate \cite{Becke-07,Tkatchenko} methods.
A simple change from Eq.~(\ref{E-ALL}) to Eq.~(\ref{my-E}) improves the theory in several
important ways: the sharp integration cutoff is obviated and consequently the $f$-sum rule is
recovered; the model system of two distant jellium spheres is properly described; accurate $C_6$
coefficients are predicted for many atoms and molecules including open-shell species.
Eq.~(\ref{my-E}) describes the asymptotic limit and has to be damped at short range.
To this end, empirical damping functions are often used (see e.g.\ Ref.~\cite{Nakai}).

The general and seamless van der Waals functional VV09 \cite{VV09} reduces to Eq.~(\ref{my-E}) in the
large separation limit. As our recent study \cite{VV09-imp} shows, VV09 performs well not only in the
asymptotic limit, but also near equilibrium intermonomer separations, provided that an adequate
exchange functional is used.

\begin{acknowledgments}
This work was supported by an NSF CAREER grant No. CHE-0547877 and a Packard Fellowship.
\end{acknowledgments}

\bibliography{C6}

\begin{thebibliography}{10}%
\makeatletter
\providecommand \@ifxundefined [1]{%
 \ifx #1\undefined \expandafter \@firstoftwo
 \else \expandafter \@secondoftwo
\fi
}%
\providecommand \@ifnum [1]{%
 \ifnum #1\expandafter \@firstoftwo
 \else \expandafter \@secondoftwo
\fi
}%
\providecommand \enquote [1]{``#1''}%
\providecommand \bibnamefont  [1]{#1}%
\providecommand \bibfnamefont [1]{#1}%
\providecommand \citenamefont [1]{#1}%
\providecommand\href[0]{\@sanitize\@href}%
\providecommand\@href[1]{\endgroup\@@startlink{#1}\endgroup\@@href}%
\providecommand\@@href[1]{#1\@@endlink}%
\providecommand \@sanitize [0]{\begingroup\catcode`\&12\catcode`\#12\relax}%
\@ifxundefined \pdfoutput {\@firstoftwo}{%
 \@ifnum{\z@=\pdfoutput}{\@firstoftwo}{\@secondoftwo}%
}{%
 \providecommand\@@startlink[1]{\leavevmode\special{html:<a href="#1">}}%
 \providecommand\@@endlink[0]{\special{html:</a>}}%
}{%
 \providecommand\@@startlink[1]{%
  \leavevmode
  \pdfstartlink
   attr{/Border[0 0 1 ]/H/I/C[0 1 1]}%
   user{/Subtype/Link/A<</Type/Action/S/URI/URI(#1)>>}%
  \relax
 }%
 \providecommand\@@endlink[0]{\pdfendlink}%
}%
\providecommand \url  [0]{\begingroup\@sanitize \@url }%
\providecommand \@url [1]{\endgroup\@href {#1}{\urlprefix}}%
\providecommand \urlprefix [0]{URL }%
\providecommand \Eprint[0]{\href }%
\@ifxundefined \urlstyle {%
  \providecommand \doi [1]{doi:\discretionary{}{}{}#1}%
}{%
  \providecommand \doi [0]{doi:\discretionary{}{}{}\begingroup
  \urlstyle{rm}\Url }%
}%
\providecommand \doibase [0]{http://dx.doi.org/}%
\providecommand \Doi[1]{\href{\doibase#1}}%
\providecommand \bibAnnote [3]{%
  \BibitemShut{#1}%
  \begin{quotation}\noindent
    \textsc{Key:}\ #2\\\textsc{Annotation:}\ #3%
  \end{quotation}%
}%
\providecommand \bibAnnoteFile [2]{%
  \IfFileExists{#2}{\bibAnnote {#1} {#2} {\input{#2}}}{}%
}%
\providecommand \typeout [0]{\immediate \write \m@ne }%
\providecommand \selectlanguage [0]{\@gobble}%
\providecommand \bibinfo [0]{\@secondoftwo}%
\providecommand \bibfield [0]{\@secondoftwo}%
\providecommand \translation [1]{[#1]}%
\providecommand \BibitemOpen[0]{}%
\providecommand \bibitemStop [0]{}%
\providecommand \bibitemNoStop [0]{.\EOS\space}%
\providecommand \EOS [0]{\spacefactor3000\relax}%
\providecommand \BibitemShut [1]{\csname bibitem#1\endcsname}%
\bibitem{VV09}%
  \BibitemOpen
  \bibfield{author}{%
  \bibinfo {author} {\bibfnamefont{O.~A.}\ \bibnamefont{Vydrov}}\ and\ \bibinfo
  {author} {\bibfnamefont{T.}~\bibnamefont{{Van Voorhis}}},\ }%
  \bibfield{journal}{%
  \bibinfo {journal} {Phys. Rev. Lett.}\ }%
  \textbf{\bibinfo {volume} {103}},\ \bibinfo {pages} {063004} (\bibinfo {year}
  {2009})%
  \bibAnnoteFile{NoStop}{VV09}%
\bibitem{vdW-DF-04}%
  \BibitemOpen
  \bibinfo {note} {M. Dion, H. Rydberg, E. Schr{\"o}der, D.~C. Langreth, and
  B.~I. Lundqvist, Phys. Rev. Lett. {\bf 92}, 246401 (2004); {\bf 95},
  109902(E) (2005)}%
  \bibAnnoteFile{NoStop}{vdW-DF-04}%
\bibitem{vdW-DF-09}%
  \BibitemOpen
  \bibfield{author}{%
  \bibinfo {author} {\bibfnamefont{O.~A.}\ \bibnamefont{Vydrov}}\ and\ \bibinfo
  {author} {\bibfnamefont{T.}~\bibnamefont{{Van Voorhis}}},\ }%
  \bibfield{journal}{%
  \bibinfo {journal} {J. Chem. Phys.}\ }%
  \textbf{\bibinfo {volume} {130}},\ \bibinfo {pages} {104105} (\bibinfo {year}
  {2009})%
  \bibAnnoteFile{NoStop}{vdW-DF-09}%
\bibitem{ALL-96}%
  \BibitemOpen
  \bibfield{author}{%
  \bibinfo {author} {\bibfnamefont{Y.}~\bibnamefont{Andersson}}, \bibinfo
  {author} {\bibfnamefont{D.~C.}\ \bibnamefont{Langreth}},\ and\ \bibinfo
  {author} {\bibfnamefont{B.~I.}\ \bibnamefont{Lundqvist}},\ }%
  \bibfield{journal}{%
  \bibinfo {journal} {Phys. Rev. Lett.}\ }%
  \textbf{\bibinfo {volume} {76}},\ \bibinfo {pages} {102} (\bibinfo {year}
  {1996})%
  \bibAnnoteFile{NoStop}{ALL-96}%
\bibitem{Dobson-96}%
  \BibitemOpen
  \bibfield{author}{%
  \bibinfo {author} {\bibfnamefont{J.~F.}\ \bibnamefont{Dobson}}\ and\ \bibinfo
  {author} {\bibfnamefont{B.~P.}\ \bibnamefont{Dinte}},\ }%
  \bibfield{journal}{%
  \bibinfo {journal} {Phys. Rev. Lett.}\ }%
  \textbf{\bibinfo {volume} {76}},\ \bibinfo {pages} {1780} (\bibinfo {year}
  {1996})%
  \bibAnnoteFile{NoStop}{Dobson-96}%
\bibitem{Kaplan}%
  \BibitemOpen
  \bibfield{author}{%
  \bibinfo {author} {\bibfnamefont{I.~G.}\ \bibnamefont{Kaplan}},\ }%
  \emph{\bibinfo {title} {Intermolecular Interactions: Physical Picture,
  Computational Methods, and Model Potentials}}\ (\bibinfo {publisher}
  {Wiley},\ \bibinfo {address} {Chichester, England},\ \bibinfo {year} {2006})%
  \bibAnnoteFile{NoStop}{Kaplan}%
\bibitem{Stephen}%
  \BibitemOpen
  \bibfield{author}{%
  \bibinfo {author} {\bibfnamefont{C.}~\bibnamefont{Mavroyannis}}\ and\
  \bibinfo {author} {\bibfnamefont{M.~J.}\ \bibnamefont{Stephen}},\ }%
  \bibfield{journal}{%
  \bibinfo {journal} {Mol. Phys.}\ }%
  \textbf{\bibinfo {volume} {5}},\ \bibinfo {pages} {629} (\bibinfo {year}
  {1962})%
  \bibAnnoteFile{NoStop}{Stephen}%
\bibitem{Book-98}%
  \BibitemOpen
  \emph{\bibinfo {title} {Electronic Density Functional Theory: Recent Progress
  and New Directions}},\ edited by\ \bibinfo {editor} {\bibfnamefont{J.~F.}\
  \bibnamefont{Dobson}}, \bibinfo {editor}
  {\bibfnamefont{G.}~\bibnamefont{Vignale}},\ and\ \bibinfo {editor}
  {\bibfnamefont{M.~P.}\ \bibnamefont{Das}}\ (\bibinfo {publisher} {Plenum},\
  \bibinfo {address} {New York},\ \bibinfo {year} {1998})%
  \bibAnnoteFile{NoStop}{Book-98}%
\bibitem{Dobson-98}%
  \BibitemOpen
  \bibinfo {note} {J.~F. Dobson, B.~P. Dinte, and J. Wang in
  Ref.~\cite{Book-98}, p. 261}%
  \bibAnnoteFile{NoStop}{Dobson-98}%
\bibitem{ALL-98}%
  \BibitemOpen
  \bibinfo {note} {Y. Andersson, E. Hult, H. Rydberg, P. Apell, B.~I.
  Lundqvist, and D.~C. Langreth in Ref.~\cite{Book-98}, p. 243}%
  \bibAnnoteFile{NoStop}{ALL-98}%
\bibitem{Rapcewicz}%
  \BibitemOpen
  \bibfield{author}{%
  \bibinfo {author} {\bibfnamefont{K.}~\bibnamefont{Rapcewicz}}\ and\ \bibinfo
  {author} {\bibfnamefont{N.~W.}\ \bibnamefont{Ashcroft}},\ }%
  \bibfield{journal}{%
  \bibinfo {journal} {Phys. Rev. B}\ }%
  \textbf{\bibinfo {volume} {44}},\ \bibinfo {pages} {4032} (\bibinfo {year}
  {1991})%
  \bibAnnoteFile{NoStop}{Rapcewicz}%
\bibitem{Mahan-book}%
  \BibitemOpen
  \bibfield{author}{%
  \bibinfo {author} {\bibfnamefont{G.~D.}\ \bibnamefont{Mahan}}\ and\ \bibinfo
  {author} {\bibfnamefont{K.~R.}\ \bibnamefont{Subbaswamy}},\ }%
  \emph{\bibinfo {title} {Local Density Theory of Polarizability}}\ (\bibinfo
  {publisher} {Plenum},\ \bibinfo {address} {New York},\ \bibinfo {year}
  {1990})%
  \bibAnnoteFile{NoStop}{Mahan-book}%
\bibitem{Nesbet-97}%
  \BibitemOpen
  \bibfield{author}{%
  \bibinfo {author} {\bibfnamefont{R.~K.}\ \bibnamefont{Nesbet}},\ }%
  \bibfield{journal}{%
  \bibinfo {journal} {Phys. Rev. A}\ }%
  \textbf{\bibinfo {volume} {56}},\ \bibinfo {pages} {2778} (\bibinfo {year}
  {1997})%
  \bibAnnoteFile{NoStop}{Nesbet-97}%
\bibitem{Nesbet-98}%
  \BibitemOpen
  \bibinfo {note} {R.~K. Nesbet in Ref.~\cite{Book-98}, p. 285}%
  \bibAnnoteFile{NoStop}{Nesbet-98}%
\bibitem{Andersson-99}%
  \BibitemOpen
  \bibfield{author}{%
  \bibinfo {author} {\bibfnamefont{Y.}~\bibnamefont{Andersson}}\ and\ \bibinfo
  {author} {\bibfnamefont{H.}~\bibnamefont{Rydberg}},\ }%
  \bibfield{journal}{%
  \bibinfo {journal} {Phys. Scr.}\ }%
  \textbf{\bibinfo {volume} {60}},\ \bibinfo {pages} {211} (\bibinfo {year}
  {1999})%
  \bibAnnoteFile{NoStop}{Andersson-99}%
\bibitem{Hirao-02}%
  \BibitemOpen
  \bibfield{author}{%
  \bibinfo {author} {\bibfnamefont{M.}~\bibnamefont{Kamiya}}, \bibinfo {author}
  {\bibfnamefont{T.}~\bibnamefont{Tsuneda}},\ and\ \bibinfo {author}
  {\bibfnamefont{K.}~\bibnamefont{Hirao}},\ }%
  \bibfield{journal}{%
  \bibinfo {journal} {J. Chem. Phys.}\ }%
  \textbf{\bibinfo {volume} {117}},\ \bibinfo {pages} {6010} (\bibinfo {year}
  {2002})%
  \bibAnnoteFile{NoStop}{Hirao-02}%
\bibitem{Hirao-05a}%
  \BibitemOpen
  \bibfield{author}{%
  \bibinfo {author} {\bibfnamefont{T.}~\bibnamefont{Sato}}, \bibinfo {author}
  {\bibfnamefont{T.}~\bibnamefont{Tsuneda}},\ and\ \bibinfo {author}
  {\bibfnamefont{K.}~\bibnamefont{Hirao}},\ }%
  \bibfield{journal}{%
  \bibinfo {journal} {J. Chem. Phys.}\ }%
  \textbf{\bibinfo {volume} {123}},\ \bibinfo {pages} {104307} (\bibinfo {year}
  {2005})%
  \bibAnnoteFile{NoStop}{Hirao-05a}%
\bibitem{Hirao-05b}%
  \BibitemOpen
  \bibfield{author}{%
  \bibinfo {author} {\bibfnamefont{T.}~\bibnamefont{Sato}}, \bibinfo {author}
  {\bibfnamefont{T.}~\bibnamefont{Tsuneda}},\ and\ \bibinfo {author}
  {\bibfnamefont{K.}~\bibnamefont{Hirao}},\ }%
  \bibfield{journal}{%
  \bibinfo {journal} {Mol. Phys.}\ }%
  \textbf{\bibinfo {volume} {103}},\ \bibinfo {pages} {1151} (\bibinfo {year}
  {2005})%
  \bibAnnoteFile{NoStop}{Hirao-05b}%
\bibitem{Hirao-07}%
  \BibitemOpen
  \bibfield{author}{%
  \bibinfo {author} {\bibfnamefont{T.}~\bibnamefont{Sato}}, \bibinfo {author}
  {\bibfnamefont{T.}~\bibnamefont{Tsuneda}},\ and\ \bibinfo {author}
  {\bibfnamefont{K.}~\bibnamefont{Hirao}},\ }%
  \bibfield{journal}{%
  \bibinfo {journal} {J. Chem. Phys.}\ }%
  \textbf{\bibinfo {volume} {126}},\ \bibinfo {pages} {234114} (\bibinfo {year}
  {2007})%
  \bibAnnoteFile{NoStop}{Hirao-07}%
\bibitem{Hirao-08}%
  \BibitemOpen
  \bibfield{author}{%
  \bibinfo {author} {\bibfnamefont{T.}~\bibnamefont{Matsui}}, \bibinfo {author}
  {\bibfnamefont{H.}~\bibnamefont{Miyachi}}, \bibinfo {author}
  {\bibfnamefont{T.}~\bibnamefont{Sato}}, \bibinfo {author}
  {\bibfnamefont{Y.}~\bibnamefont{Shigeta}},\ and\ \bibinfo {author}
  {\bibfnamefont{K.}~\bibnamefont{Hirao}},\ }%
  \bibfield{journal}{%
  \bibinfo {journal} {J. Phys. Chem. B}\ }%
  \textbf{\bibinfo {volume} {112}},\ \bibinfo {pages} {16960} (\bibinfo {year}
  {2008})%
  \bibAnnoteFile{NoStop}{Hirao-08}%
\bibitem{Hirao-09a}%
  \BibitemOpen
  \bibfield{author}{%
  \bibinfo {author} {\bibfnamefont{T.}~\bibnamefont{Matsui}}, \bibinfo {author}
  {\bibfnamefont{T.}~\bibnamefont{Sato}}, \bibinfo {author}
  {\bibfnamefont{Y.}~\bibnamefont{Shigeta}},\ and\ \bibinfo {author}
  {\bibfnamefont{K.}~\bibnamefont{Hirao}},\ }%
  \bibfield{journal}{%
  \bibinfo {journal} {Chem. Phys. Lett.}\ }%
  \textbf{\bibinfo {volume} {478}},\ \bibinfo {pages} {238} (\bibinfo {year}
  {2009})%
  \bibAnnoteFile{NoStop}{Hirao-09a}%
\bibitem{Hirao-09b}%
  \BibitemOpen
  \bibfield{author}{%
  \bibinfo {author} {\bibfnamefont{T.}~\bibnamefont{Matsui}}, \bibinfo {author}
  {\bibfnamefont{H.}~\bibnamefont{Miyachi}}, \bibinfo {author}
  {\bibfnamefont{Y.}~\bibnamefont{Nakanishi}}, \bibinfo {author}
  {\bibfnamefont{Y.}~\bibnamefont{Shigeta}}, \bibinfo {author}
  {\bibfnamefont{T.}~\bibnamefont{Sato}}, \bibinfo {author}
  {\bibfnamefont{Y.}~\bibnamefont{Kitagawa}}, \bibinfo {author}
  {\bibfnamefont{M.}~\bibnamefont{Okumura}},\ and\ \bibinfo {author}
  {\bibfnamefont{K.}~\bibnamefont{Hirao}},\ }%
  \bibfield{journal}{%
  \bibinfo {journal} {J. Phys. Chem. B}\ }%
  \textbf{\bibinfo {volume} {113}},\ \bibinfo {pages} {12790} (\bibinfo {year}
  {2009})%
  \bibAnnoteFile{NoStop}{Hirao-09b}%
\bibitem{Hirao-10}%
  \BibitemOpen
  \bibfield{author}{%
  \bibinfo {author} {\bibfnamefont{H.}~\bibnamefont{Miyachi}}, \bibinfo
  {author} {\bibfnamefont{T.}~\bibnamefont{Matsui}}, \bibinfo {author}
  {\bibfnamefont{Y.}~\bibnamefont{Shigeta}},\ and\ \bibinfo {author}
  {\bibfnamefont{K.}~\bibnamefont{Hirao}},\ }%
  \bibfield{journal}{%
  \bibinfo {journal} {Phys. Chem. Chem. Phys.}\ }%
  \textbf{\bibinfo {volume} {12}},\ \bibinfo {pages} {909} (\bibinfo {year}
  {2010})%
  \bibAnnoteFile{NoStop}{Hirao-10}%
\bibitem{Silvestrelli-08}%
  \BibitemOpen
  \bibfield{author}{%
  \bibinfo {author} {\bibfnamefont{P.~L.}\ \bibnamefont{Silvestrelli}},\ }%
  \bibfield{journal}{%
  \bibinfo {journal} {Phys. Rev. Lett.}\ }%
  \textbf{\bibinfo {volume} {100}},\ \bibinfo {pages} {053002} (\bibinfo {year}
  {2008})%
  \bibAnnoteFile{NoStop}{Silvestrelli-08}%
\bibitem{Silvestrelli-09a}%
  \BibitemOpen
  \bibfield{author}{%
  \bibinfo {author} {\bibfnamefont{P.~L.}\ \bibnamefont{Silvestrelli}},
  \bibinfo {author} {\bibfnamefont{K.}~\bibnamefont{Benyahia}}, \bibinfo
  {author} {\bibfnamefont{S.}~\bibnamefont{Grubisi\^c}}, \bibinfo {author}
  {\bibfnamefont{F.}~\bibnamefont{Ancilotto}},\ and\ \bibinfo {author}
  {\bibfnamefont{F.}~\bibnamefont{Toigo}},\ }%
  \bibfield{journal}{%
  \bibinfo {journal} {J. Chem. Phys.}\ }%
  \textbf{\bibinfo {volume} {130}},\ \bibinfo {pages} {074702} (\bibinfo {year}
  {2009})%
  \bibAnnoteFile{NoStop}{Silvestrelli-09a}%
\bibitem{Silvestrelli-09b}%
  \BibitemOpen
  \bibfield{author}{%
  \bibinfo {author} {\bibfnamefont{P.~L.}\ \bibnamefont{Silvestrelli}},\ }%
  \bibfield{journal}{%
  \bibinfo {journal} {J. Phys. Chem. A}\ }%
  \textbf{\bibinfo {volume} {113}},\ \bibinfo {pages} {5224} (\bibinfo {year}
  {2009})%
  \bibAnnoteFile{NoStop}{Silvestrelli-09b}%
\bibitem{Silvestrelli-09c}%
  \BibitemOpen
  \bibfield{author}{%
  \bibinfo {author} {\bibfnamefont{P.~L.}\ \bibnamefont{Silvestrelli}},\ }%
  \bibfield{journal}{%
  \bibinfo {journal} {Chem. Phys. Lett.}\ }%
  \textbf{\bibinfo {volume} {475}},\ \bibinfo {pages} {285} (\bibinfo {year}
  {2009})%
  \bibAnnoteFile{NoStop}{Silvestrelli-09c}%
\bibitem{Silvestrelli-09d}%
  \BibitemOpen
  \bibfield{author}{%
  \bibinfo {author} {\bibfnamefont{P.~L.}\ \bibnamefont{Silvestrelli}},
  \bibinfo {author} {\bibfnamefont{F.}~\bibnamefont{Toigo}},\ and\ \bibinfo
  {author} {\bibfnamefont{F.}~\bibnamefont{Ancilotto}},\ }%
  \bibfield{journal}{%
  \bibinfo {journal} {J. Phys. Chem. C}\ }%
  \textbf{\bibinfo {volume} {113}},\ \bibinfo {pages} {17124} (\bibinfo {year}
  {2009})%
  \bibAnnoteFile{NoStop}{Silvestrelli-09d}%
\bibitem{Cremer-09}%
  \BibitemOpen
  \bibfield{author}{%
  \bibinfo {author} {\bibfnamefont{J.}~\bibnamefont{Gr{\" a}fenstein}}\ and\
  \bibinfo {author} {\bibfnamefont{D.}~\bibnamefont{Cremer}},\ }%
  \bibfield{journal}{%
  \bibinfo {journal} {J. Chem. Phys.}\ }%
  \textbf{\bibinfo {volume} {130}},\ \bibinfo {pages} {124105} (\bibinfo {year}
  {2009})%
  \bibAnnoteFile{NoStop}{Cremer-09}%
\bibitem{Hagiwara-09}%
  \BibitemOpen
  \bibfield{author}{%
  \bibinfo {author} {\bibfnamefont{Y.}~\bibnamefont{Hagiwara}}\ and\ \bibinfo
  {author} {\bibfnamefont{M.}~\bibnamefont{Tateno}},\ }%
  \bibfield{journal}{%
  \bibinfo {journal} {J. Phys.: Condens. Matter}\ }%
  \textbf{\bibinfo {volume} {21}},\ \bibinfo {pages} {245103} (\bibinfo {year}
  {2009})%
  \bibAnnoteFile{NoStop}{Hagiwara-09}%
\bibitem{Rotenberg-10}%
  \BibitemOpen
  \bibfield{author}{%
  \bibinfo {author} {\bibfnamefont{B.}~\bibnamefont{Rotenberg}}, \bibinfo
  {author} {\bibfnamefont{M.}~\bibnamefont{Salanne}}, \bibinfo {author}
  {\bibfnamefont{C.}~\bibnamefont{Simon}},\ and\ \bibinfo {author}
  {\bibfnamefont{R.}~\bibnamefont{Vuilleumier}},\ }%
  \bibfield{journal}{%
  \bibinfo {journal} {Phys. Rev. Lett.}\ }%
  \textbf{\bibinfo {volume} {104}},\ \bibinfo {pages} {138301} (\bibinfo {year}
  {2010})%
  \bibAnnoteFile{NoStop}{Rotenberg-10}%
\bibitem{spheres}%
  \BibitemOpen
  \bibfield{author}{%
  \bibinfo {author} {\bibfnamefont{A.~A.}\ \bibnamefont{Lucas}}, \bibinfo
  {author} {\bibfnamefont{A.}~\bibnamefont{Ronveaux}}, \bibinfo {author}
  {\bibfnamefont{M.}~\bibnamefont{Schmeits}},\ and\ \bibinfo {author}
  {\bibfnamefont{F.}~\bibnamefont{Delanaye}},\ }%
  \bibfield{journal}{%
  \bibinfo {journal} {Phys. Rev. B}\ }%
  \textbf{\bibinfo {volume} {12}},\ \bibinfo {pages} {5372} (\bibinfo {year}
  {1975})%
  \bibAnnoteFile{NoStop}{spheres}%
\bibitem{vdW-DF-10}%
  \BibitemOpen
  \bibfield{author}{%
  \bibinfo {author} {\bibfnamefont{K.}~\bibnamefont{Lee}}, \bibinfo {author}
  {\bibfnamefont{{\'E}.~D.}\ \bibnamefont{Murray}}, \bibinfo {author}
  {\bibfnamefont{L.}~\bibnamefont{Kong}}, \bibinfo {author}
  {\bibfnamefont{B.~I.}\ \bibnamefont{Lundqvist}},\ and\ \bibinfo {author}
  {\bibfnamefont{D.~C.}\ \bibnamefont{Langreth}},\ }%
  \Eprint{http://arxiv.org/abs/1003.5255}{arXiv:1003.5255}%
  \bibAnnoteFile{NoStop}{vdW-DF-10}%
\bibitem{Levine}%
  \BibitemOpen
  \bibfield{author}{%
  \bibinfo {author} {\bibfnamefont{Z.~H.}\ \bibnamefont{Levine}}\ and\ \bibinfo
  {author} {\bibfnamefont{S.~G.}\ \bibnamefont{Louie}},\ }%
  \bibfield{journal}{%
  \bibinfo {journal} {Phys. Rev. B}\ }%
  \textbf{\bibinfo {volume} {25}},\ \bibinfo {pages} {6310} (\bibinfo {year}
  {1982})%
  \bibAnnoteFile{NoStop}{Levine}%
\bibitem{Levy-84}%
  \BibitemOpen
  \bibfield{author}{%
  \bibinfo {author} {\bibfnamefont{M.}~\bibnamefont{Levy}}, \bibinfo {author}
  {\bibfnamefont{J.~P.}\ \bibnamefont{Perdew}},\ and\ \bibinfo {author}
  {\bibfnamefont{V.}~\bibnamefont{Sahni}},\ }%
  \bibfield{journal}{%
  \bibinfo {journal} {Phys. Rev. A}\ }%
  \textbf{\bibinfo {volume} {30}},\ \bibinfo {pages} {2745} (\bibinfo {year}
  {1984})%
  \bibAnnoteFile{NoStop}{Levy-84}%
\bibitem{Savin-99}%
  \BibitemOpen
  \bibfield{author}{%
  \bibinfo {author} {\bibfnamefont{C.}~\bibnamefont{Gutle}}, \bibinfo {author}
  {\bibfnamefont{A.}~\bibnamefont{Savin}}, \bibinfo {author}
  {\bibfnamefont{J.~B.}\ \bibnamefont{Krieger}},\ and\ \bibinfo {author}
  {\bibfnamefont{J.}~\bibnamefont{Chen}},\ }%
  \bibfield{journal}{%
  \bibinfo {journal} {Int. J. Quantum Chem.}\ }%
  \textbf{\bibinfo {volume} {75}},\ \bibinfo {pages} {885} (\bibinfo {year}
  {1999})%
  \bibAnnoteFile{NoStop}{Savin-99}%
\bibitem{Savin-07}%
  \BibitemOpen
  \bibfield{author}{%
  \bibinfo {author} {\bibfnamefont{C.}~\bibnamefont{Gutl\'e}}\ and\ \bibinfo
  {author} {\bibfnamefont{A.}~\bibnamefont{Savin}},\ }%
  \bibfield{journal}{%
  \bibinfo {journal} {Phys. Rev. A}\ }%
  \textbf{\bibinfo {volume} {75}},\ \bibinfo {pages} {032519} (\bibinfo {year}
  {2007})%
  \bibAnnoteFile{NoStop}{Savin-07}%
\bibitem{LiH}%
  \BibitemOpen
  \bibfield{author}{%
  \bibinfo {author} {\bibfnamefont{G.~L.}\ \bibnamefont{Bendazzoli}}, \bibinfo
  {author} {\bibfnamefont{V.}~\bibnamefont{Magnasco}}, \bibinfo {author}
  {\bibfnamefont{G.}~\bibnamefont{Figari}},\ and\ \bibinfo {author}
  {\bibfnamefont{M.}~\bibnamefont{Rui}},\ }%
  \bibfield{journal}{%
  \bibinfo {journal} {Chem. Phys. Lett.}\ }%
  \textbf{\bibinfo {volume} {330}},\ \bibinfo {pages} {146} (\bibinfo {year}
  {2000})%
  \bibAnnoteFile{NoStop}{LiH}%
\bibitem{Derevianko-10}%
  \BibitemOpen
  \bibfield{author}{%
  \bibinfo {author} {\bibfnamefont{A.}~\bibnamefont{Derevianko}}, \bibinfo
  {author} {\bibfnamefont{S.~G.}\ \bibnamefont{Porsev}},\ and\ \bibinfo
  {author} {\bibfnamefont{J.~F.}\ \bibnamefont{Babb}},\ }%
  \bibfield{journal}{%
  \bibinfo {journal} {At. Data. Nucl. Data Tables}\ }%
  \textbf{\bibinfo {volume} {96}},\ \bibinfo {pages} {323} (\bibinfo {year}
  {2010})%
  \bibAnnoteFile{NoStop}{Derevianko-10}%
\bibitem{BeH2}%
  \BibitemOpen
  \bibfield{author}{%
  \bibinfo {author} {\bibfnamefont{G.~L.}\ \bibnamefont{Bendazzoli}}, \bibinfo
  {author} {\bibfnamefont{A.}~\bibnamefont{Monari}}, \bibinfo {author}
  {\bibfnamefont{G.}~\bibnamefont{Figari}}, \bibinfo {author}
  {\bibfnamefont{M.}~\bibnamefont{Rui}}, \bibinfo {author}
  {\bibfnamefont{C.}~\bibnamefont{Costa}},\ and\ \bibinfo {author}
  {\bibfnamefont{V.}~\bibnamefont{Magnasco}},\ }%
  \bibfield{journal}{%
  \bibinfo {journal} {Chem. Phys. Lett.}\ }%
  \textbf{\bibinfo {volume} {414}},\ \bibinfo {pages} {51} (\bibinfo {year}
  {2005})%
  \bibAnnoteFile{NoStop}{BeH2}%
\bibitem{BH}%
  \BibitemOpen
  \bibfield{author}{%
  \bibinfo {author} {\bibfnamefont{G.~L.}\ \bibnamefont{Bendazzoli}}, \bibinfo
  {author} {\bibfnamefont{A.}~\bibnamefont{Monari}}, \bibinfo {author}
  {\bibfnamefont{G.}~\bibnamefont{Figari}}, \bibinfo {author}
  {\bibfnamefont{M.}~\bibnamefont{Rui}}, \bibinfo {author}
  {\bibfnamefont{C.}~\bibnamefont{Costa}},\ and\ \bibinfo {author}
  {\bibfnamefont{V.}~\bibnamefont{Magnasco}},\ }%
  \bibfield{journal}{%
  \bibinfo {journal} {Chem. Phys. Lett.}\ }%
  \textbf{\bibinfo {volume} {450}},\ \bibinfo {pages} {396} (\bibinfo {year}
  {2008})%
  \bibAnnoteFile{NoStop}{BH}%
\bibitem{LC-wPBE08}%
  \BibitemOpen
  \bibfield{author}{%
  \bibinfo {author} {\bibfnamefont{E.}~\bibnamefont{Weintraub}}, \bibinfo
  {author} {\bibfnamefont{T.~M.}\ \bibnamefont{Henderson}},\ and\ \bibinfo
  {author} {\bibfnamefont{G.~E.}\ \bibnamefont{Scuseria}},\ }%
  \bibfield{journal}{%
  \bibinfo {journal} {J. Chem. Theory Comput.}\ }%
  \textbf{\bibinfo {volume} {5}},\ \bibinfo {pages} {754} (\bibinfo {year}
  {2009})%
  \bibAnnoteFile{NoStop}{LC-wPBE08}%
\bibitem{CRC}%
  \BibitemOpen
  \emph{\bibinfo {title} {CRC Handbook of Chemistry and Physics}},\ \bibinfo
  {edition} {90th}\ ed.,\ edited by\ \bibinfo {editor} {\bibfnamefont{D.~R.}\
  \bibnamefont{Lide}}\ (\bibinfo {publisher} {CRC Press},\ \bibinfo {address}
  {Boca Raton, FL},\ \bibinfo {year} {2009})%
  \bibAnnoteFile{NoStop}{CRC}%
\bibitem{Kumar-10}%
  \BibitemOpen
  \bibfield{author}{%
  \bibinfo {author} {\bibfnamefont{A.}~\bibnamefont{Kumar}}\ and\ \bibinfo
  {author} {\bibfnamefont{A.~J.}\ \bibnamefont{Thakkar}},\ }%
  \bibfield{journal}{%
  \bibinfo {journal} {J. Chem. Phys.}\ }%
  \textbf{\bibinfo {volume} {132}},\ \bibinfo {pages} {074301} (\bibinfo {year}
  {2010})%
  \bibAnnoteFile{NoStop}{Kumar-10}%
\bibitem{Chu-JCP-04}%
  \BibitemOpen
  \bibfield{author}{%
  \bibinfo {author} {\bibfnamefont{X.}~\bibnamefont{Chu}}\ and\ \bibinfo
  {author} {\bibfnamefont{A.}~\bibnamefont{Dalgarno}},\ }%
  \bibfield{journal}{%
  \bibinfo {journal} {J. Chem. Phys.}\ }%
  \textbf{\bibinfo {volume} {121}},\ \bibinfo {pages} {4083} (\bibinfo {year}
  {2004})%
  \bibAnnoteFile{NoStop}{Chu-JCP-04}%
\bibitem{Meath-90-IntJQC}%
  \BibitemOpen
  \bibfield{author}{%
  \bibinfo {author} {\bibfnamefont{W.~J.}\ \bibnamefont{Meath}}\ and\ \bibinfo
  {author} {\bibfnamefont{A.}~\bibnamefont{Kumar}},\ }%
  \bibfield{journal}{%
  \bibinfo {journal} {Int. J. Quantum Chem. Symp.}\ }%
  \textbf{\bibinfo {volume} {24}},\ \bibinfo {pages} {501} (\bibinfo {year}
  {1990})%
  \bibAnnoteFile{NoStop}{Meath-90-IntJQC}%
\bibitem{Meath-02-MolPhys}%
  \BibitemOpen
  \bibfield{author}{%
  \bibinfo {author} {\bibfnamefont{M.}~\bibnamefont{Kumar}}, \bibinfo {author}
  {\bibfnamefont{A.}~\bibnamefont{Kumar}},\ and\ \bibinfo {author}
  {\bibfnamefont{W.~J.}\ \bibnamefont{Meath}},\ }%
  \bibfield{journal}{%
  \bibinfo {journal} {Mol. Phys.}\ }%
  \textbf{\bibinfo {volume} {100}},\ \bibinfo {pages} {3271} (\bibinfo {year}
  {2002})%
  \bibAnnoteFile{NoStop}{Meath-02-MolPhys}%
\bibitem{Meath-85-MolPhys}%
  \BibitemOpen
  \bibfield{author}{%
  \bibinfo {author} {\bibfnamefont{A.}~\bibnamefont{Kumar}}\ and\ \bibinfo
  {author} {\bibfnamefont{W.~J.}\ \bibnamefont{Meath}},\ }%
  \bibfield{journal}{%
  \bibinfo {journal} {Mol. Phys.}\ }%
  \textbf{\bibinfo {volume} {54}},\ \bibinfo {pages} {823} (\bibinfo {year}
  {1985})%
  \bibAnnoteFile{NoStop}{Meath-85-MolPhys}%
\bibitem{Meath-82-CP}%
  \BibitemOpen
  \bibfield{author}{%
  \bibinfo {author} {\bibfnamefont{B.~L.}\ \bibnamefont{Jhanwar}}\ and\
  \bibinfo {author} {\bibfnamefont{W.~J.}\ \bibnamefont{Meath}},\ }%
  \bibfield{journal}{%
  \bibinfo {journal} {Chem. Phys.}\ }%
  \textbf{\bibinfo {volume} {67}},\ \bibinfo {pages} {185} (\bibinfo {year}
  {1982})%
  \bibAnnoteFile{NoStop}{Meath-82-CP}%
\bibitem{Meath-84-CP}%
  \BibitemOpen
  \bibfield{author}{%
  \bibinfo {author} {\bibfnamefont{A.}~\bibnamefont{Kumar}}\ and\ \bibinfo
  {author} {\bibfnamefont{W.~J.}\ \bibnamefont{Meath}},\ }%
  \bibfield{journal}{%
  \bibinfo {journal} {Chem. Phys.}\ }%
  \textbf{\bibinfo {volume} {91}},\ \bibinfo {pages} {411} (\bibinfo {year}
  {1984})%
  \bibAnnoteFile{NoStop}{Meath-84-CP}%
\bibitem{Meath-78-JCP}%
  \BibitemOpen
  \bibfield{author}{%
  \bibinfo {author} {\bibfnamefont{D.~J.}\ \bibnamefont{Margoliash}}\ and\
  \bibinfo {author} {\bibfnamefont{W.~J.}\ \bibnamefont{Meath}},\ }%
  \bibfield{journal}{%
  \bibinfo {journal} {J. Chem. Phys.}\ }%
  \textbf{\bibinfo {volume} {68}},\ \bibinfo {pages} {1426} (\bibinfo {year}
  {1978})%
  \bibAnnoteFile{NoStop}{Meath-78-JCP}%
\bibitem{Meath-80-CP}%
  \BibitemOpen
  \bibfield{author}{%
  \bibinfo {author} {\bibfnamefont{G.~F.}\ \bibnamefont{Thomas}}, \bibinfo
  {author} {\bibfnamefont{F.}~\bibnamefont{Mulder}},\ and\ \bibinfo {author}
  {\bibfnamefont{W.~J.}\ \bibnamefont{Meath}},\ }%
  \bibfield{journal}{%
  \bibinfo {journal} {Chem. Phys.}\ }%
  \textbf{\bibinfo {volume} {54}},\ \bibinfo {pages} {45} (\bibinfo {year}
  {1980})%
  \bibAnnoteFile{NoStop}{Meath-80-CP}%
\bibitem{Meath-02-THEO}%
  \BibitemOpen
  \bibfield{author}{%
  \bibinfo {author} {\bibfnamefont{A.}~\bibnamefont{Kumar}},\ }%
  \bibfield{journal}{%
  \bibinfo {journal} {J. Mol. Struct. THEOCHEM}\ }%
  \textbf{\bibinfo {volume} {591}},\ \bibinfo {pages} {91} (\bibinfo {year}
  {2002})%
  \bibAnnoteFile{NoStop}{Meath-02-THEO}%
\bibitem{Meath-03-CP}%
  \BibitemOpen
  \bibfield{author}{%
  \bibinfo {author} {\bibfnamefont{A.}~\bibnamefont{Kumar}}, \bibinfo {author}
  {\bibfnamefont{M.}~\bibnamefont{Kumar}},\ and\ \bibinfo {author}
  {\bibfnamefont{W.~J.}\ \bibnamefont{Meath}},\ }%
  \bibfield{journal}{%
  \bibinfo {journal} {Chem. Phys.}\ }%
  \textbf{\bibinfo {volume} {286}},\ \bibinfo {pages} {227} (\bibinfo {year}
  {2003})%
  \bibAnnoteFile{NoStop}{Meath-03-CP}%
\bibitem{Meath-03-MolPhys}%
  \BibitemOpen
  \bibfield{author}{%
  \bibinfo {author} {\bibfnamefont{A.}~\bibnamefont{Kumar}}, \bibinfo {author}
  {\bibfnamefont{M.}~\bibnamefont{Kumar}},\ and\ \bibinfo {author}
  {\bibfnamefont{W.~J.}\ \bibnamefont{Meath}},\ }%
  \bibfield{journal}{%
  \bibinfo {journal} {Mol. Phys.}\ }%
  \textbf{\bibinfo {volume} {101}},\ \bibinfo {pages} {1535} (\bibinfo {year}
  {2003})%
  \bibAnnoteFile{NoStop}{Meath-03-MolPhys}%
\bibitem{Meath-88-CanJChem}%
  \BibitemOpen
  \bibfield{author}{%
  \bibinfo {author} {\bibfnamefont{R.~J.}\ \bibnamefont{Pazur}}, \bibinfo
  {author} {\bibfnamefont{A.}~\bibnamefont{Kumar}}, \bibinfo {author}
  {\bibfnamefont{R.~A.}\ \bibnamefont{Thuraisingham}},\ and\ \bibinfo {author}
  {\bibfnamefont{W.~J.}\ \bibnamefont{Meath}},\ }%
  \bibfield{journal}{%
  \bibinfo {journal} {Can. J. Chem.}\ }%
  \textbf{\bibinfo {volume} {66}},\ \bibinfo {pages} {615} (\bibinfo {year}
  {1988})%
  \bibAnnoteFile{NoStop}{Meath-88-CanJChem}%
\bibitem{Meath-85-JCP}%
  \BibitemOpen
  \bibfield{author}{%
  \bibinfo {author} {\bibfnamefont{A.}~\bibnamefont{Kumar}}, \bibinfo {author}
  {\bibfnamefont{G.~R.~G.}\ \bibnamefont{Fairley}},\ and\ \bibinfo {author}
  {\bibfnamefont{W.~J.}\ \bibnamefont{Meath}},\ }%
  \bibfield{journal}{%
  \bibinfo {journal} {J. Chem. Phys.}\ }%
  \textbf{\bibinfo {volume} {83}},\ \bibinfo {pages} {70} (\bibinfo {year}
  {1985})%
  \bibAnnoteFile{NoStop}{Meath-85-JCP}%
\bibitem{Meath-92-MolPhys}%
  \BibitemOpen
  \bibfield{author}{%
  \bibinfo {author} {\bibfnamefont{A.}~\bibnamefont{Kumar}}\ and\ \bibinfo
  {author} {\bibfnamefont{W.~J.}\ \bibnamefont{Meath}},\ }%
  \bibfield{journal}{%
  \bibinfo {journal} {Mol. Phys.}\ }%
  \textbf{\bibinfo {volume} {75}},\ \bibinfo {pages} {311} (\bibinfo {year}
  {1992})%
  \bibAnnoteFile{NoStop}{Meath-92-MolPhys}%
\bibitem{Meath-07-CanJChem}%
  \BibitemOpen
  \bibfield{author}{%
  \bibinfo {author} {\bibfnamefont{A.}~\bibnamefont{Kumar}}, \bibinfo {author}
  {\bibfnamefont{B.~L.}\ \bibnamefont{Jhanwar}},\ and\ \bibinfo {author}
  {\bibfnamefont{W.}~\bibnamefont{Meath}},\ }%
  \bibfield{journal}{%
  \bibinfo {journal} {Can. J. Chem.}\ }%
  \textbf{\bibinfo {volume} {85}},\ \bibinfo {pages} {724} (\bibinfo {year}
  {2007})%
  \bibAnnoteFile{NoStop}{Meath-07-CanJChem}%
\bibitem{Meath-05-CzechCC}%
  \BibitemOpen
  \bibfield{author}{%
  \bibinfo {author} {\bibfnamefont{A.}~\bibnamefont{Kumar}}, \bibinfo {author}
  {\bibfnamefont{B.~L.}\ \bibnamefont{Jhanwar}},\ and\ \bibinfo {author}
  {\bibfnamefont{W.~J.}\ \bibnamefont{Meath}},\ }%
  \bibfield{journal}{%
  \bibinfo {journal} {Collect. Czech. Chem. Commun.}\ }%
  \textbf{\bibinfo {volume} {70}},\ \bibinfo {pages} {1196} (\bibinfo {year}
  {2005})%
  \bibAnnoteFile{NoStop}{Meath-05-CzechCC}%
\bibitem{Meath-08-MolPhys}%
  \BibitemOpen
  \bibfield{author}{%
  \bibinfo {author} {\bibfnamefont{A.}~\bibnamefont{Kumar}}\ and\ \bibinfo
  {author} {\bibfnamefont{W.~J.}\ \bibnamefont{Meath}},\ }%
  \bibfield{journal}{%
  \bibinfo {journal} {Mol. Phys.}\ }%
  \textbf{\bibinfo {volume} {106}},\ \bibinfo {pages} {1531} (\bibinfo {year}
  {2008})%
  \bibAnnoteFile{NoStop}{Meath-08-MolPhys}%
\bibitem{Nakai}%
  \BibitemOpen
  \bibfield{author}{%
  \bibinfo {author} {\bibfnamefont{T.}~\bibnamefont{Sato}}\ and\ \bibinfo
  {author} {\bibfnamefont{H.}~\bibnamefont{Nakai}},\ }%
  \bibfield{journal}{%
  \bibinfo {journal} {J. Chem. Phys.}\ }%
  \textbf{\bibinfo {volume} {131}},\ \bibinfo {pages} {224104} (\bibinfo {year}
  {2009})%
  \bibAnnoteFile{NoStop}{Nakai}%
\bibitem{Becke-07}%
  \BibitemOpen
  \bibfield{author}{%
  \bibinfo {author} {\bibfnamefont{A.~D.}\ \bibnamefont{Becke}}\ and\ \bibinfo
  {author} {\bibfnamefont{E.~R.}\ \bibnamefont{Johnson}},\ }%
  \bibfield{journal}{%
  \bibinfo {journal} {J. Chem. Phys.}\ }%
  \textbf{\bibinfo {volume} {127}},\ \bibinfo {pages} {154108} (\bibinfo {year}
  {2007})%
  \bibAnnoteFile{NoStop}{Becke-07}%
\bibitem{Tkatchenko}%
  \BibitemOpen
  \bibfield{author}{%
  \bibinfo {author} {\bibfnamefont{A.}~\bibnamefont{Tkatchenko}}\ and\ \bibinfo
  {author} {\bibfnamefont{M.}~\bibnamefont{Scheffler}},\ }%
  \bibfield{journal}{%
  \bibinfo {journal} {Phys. Rev. Lett.}\ }%
  \textbf{\bibinfo {volume} {102}},\ \bibinfo {pages} {073005} (\bibinfo {year}
  {2009})%
  \bibAnnoteFile{NoStop}{Tkatchenko}%
\bibitem{VV09-imp}%
  \BibitemOpen
  \bibfield{author}{%
  \bibinfo {author} {\bibfnamefont{O.~A.}\ \bibnamefont{Vydrov}}\ and\ \bibinfo
  {author} {\bibfnamefont{T.}~\bibnamefont{{Van Voorhis}}},\ }%
  \bibfield{journal}{%
  \bibinfo {journal} {J. Chem. Phys.}\ }%
  \textbf{\bibinfo {volume} {132}},\ \bibinfo {pages} {164113} (\bibinfo {year}
  {2010})%
  \bibAnnoteFile{NoStop}{VV09-imp}%
\end{thebibliography}%

\end{document}